\documentclass[journal,12pt,onecolumn]{IEEEtran}

\usepackage[pdftex]{graphicx}
\usepackage{fullpage}
\usepackage{cite}

\usepackage[cmex10]{amsmath}
\usepackage{algorithmic}
\usepackage{mdwmath}
\usepackage{mdwtab}
\usepackage{amssymb}
\usepackage[tight,footnotesize]{subfigure}

\newtheorem{theorem}{Theorem}

\newenvironment{proof}[1][Proof]{\begin{trivlist}
\item[\hskip \labelsep {\bfseries #1}]}{\end{trivlist}}

\newcommand{\qed}{\nobreak \ifvmode \relax \else
      \ifdim\lastskip<1.5em \hskip-\lastskip
      \hskip1.5em plus0em minus0.5em \fi \nobreak
      \vrule height0.75em width0.5em depth0.25em\fi}

\begin{document}
\title{The Induced Bounded-Degree Subgraph Problem and 
Stream Control\\ in MIMO Networks}

\author{\IEEEauthorblockN{Harish Sethu}\\
\IEEEauthorblockA{Email: {\tt sethu@drexel.edu}}\\
Department of Electrical and Computer Engineering\\
Drexel University, Philadelphia, PA 19104-2875
}

\maketitle

~\vskip 1in
\begin{abstract}
In this report, we consider maximal solutions to the induced
bounded-degree subgraph problem and relate it to issues
concerning stream control in multiple-input multiple-output (MIMO)
networks. We present a new distributed algorithm that completes in
logarithmic time with high probability and is guaranteed to complete
in linear time. We conclude the report with simulation results that
address the effectiveness of stream control and the relative impact of
receiver overloading and flexible interference suppression.
\end{abstract}
\footnotetext[1]{This work was supported in part by NSF Award CNS-0322797.} 

\IEEEpeerreviewmaketitle
\thispagestyle{empty}

\newpage
\section{Introduction}
\label{sec:intro}

Multi-hop wireless sensor networks and mobile ad hoc networks
have been suggested for a variety of applications in both military and
civilian environments. In such networks, the aggregate
traffic-carrying capacity is determined by the number of concurrent
transmissions that can be supported without interference from other
transmissions. If we consider each transmission link in a
network as a vertex in a graph with an edge connecting each pair of
vertices corresponding to interfering links, the problem of finding the maximum
aggregate capacity reduces to the problem of finding the largest
independent set in the graph \cite{Pel2000,BalBar2004}. In this
report, we pose a generalized version of the independent set problem
inspired by the technology of multiple-input multiple-output (MIMO)
enabled by smart antennas \cite{HayMoh2005}.

\begin{figure}[!t]
\begin{center}
\includegraphics[width=5.0in]{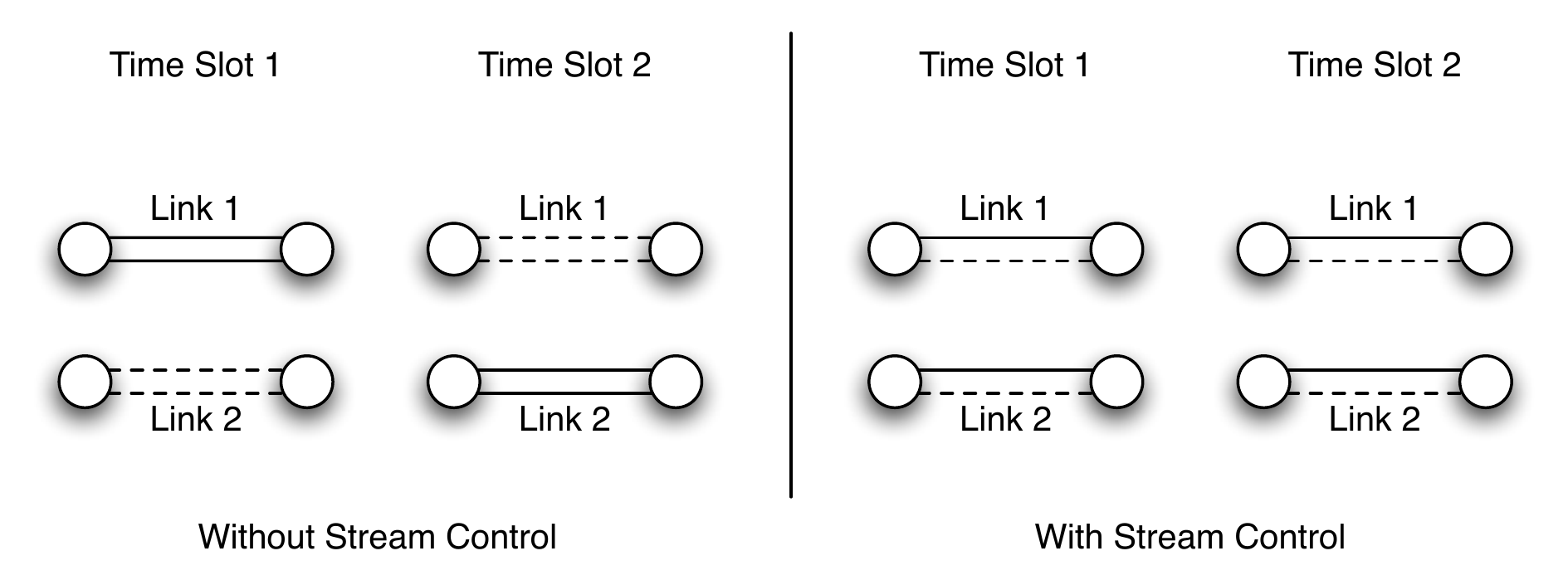}
\caption{Two interfering MIMO links, each comprised of two streams, can achieve a higher data 
rate if both links transmit simultaneously on their stronger
streams instead of taking turns and transmitting during different
slots on all streams. Solid line indicates an active stream while a
dashed line indicates an unused stream.} 
\label{fig:sc}
\end{center}
\end{figure}

In a MIMO link, multiple antenna elements at both ends of a communication
link allow simultaneous transmission of multiple data streams, thus
achieving a higher capacity without increased bandwidth or power
requirements. With $q$ degrees of freedom (number of antenna
elements), a MIMO receiver can isolate and decipher up to $q$ streams
that reach the receiver. A MIMO link, however, may choose only the
strongest streams (channel modes) and transmit on fewer than $q$
streams. Such {\em stream control}, illustrated in Fig.\,\,\ref{fig:sc}, is known to allow multiple 
interfering links operating simultaneously to achieve a better overall
throughput than if each link transmits on all 
available streams \cite{DemIng2003,BluWin2002}. As identified in
\cite{SunSiv2004}, there are at least two aspects 
related to stream control that influence, in opposite directions, the
potential capacity of a MIMO network. The phenomenon of {\em receiver
  overloading} offsets the gains from stream control because an active
receiver in a contention zone reduces the number of
potential transmitters in all the contention zones that it belongs
to. This can reduce the ability to exploit spatial diversity since more links are active when stream
control is employed and therefore, more active receivers may be
overloaded. On the other hand, stream control allows {\em flexible 
  interference suppression} because the number of degrees of freedom
required to suppress interference at a node may be smaller than the
number of interfering streams. For example, it may be possible to
decode two desired streams using up two degrees of freedom while
suppressing two other undesirable interfering streams using up only
one additional degree of freedom, thus using a total of only three
(instead of four) degrees of freedom. 

In this report, we seek a graph-theoretic approach to gain some
insight into the relative influence of receiver overloading and
flexible interference suppression on the effectiveness of stream
control in a MIMO network. We relate the issue to the problem of
finding a maximal induced bounded-degree subgraph of the contention
graph of a MIMO network.

\section{Problem Statement}
\label{sec:problem}

\begin{figure}[!t]
\begin{center}
    \subfigure[{}]{
        \includegraphics[width=3.0in]{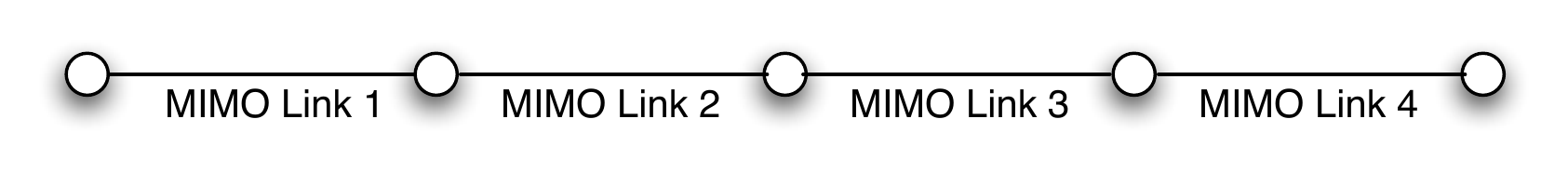}
        \label{fig:example-MIMO-network}
        }\\
    \subfigure[{}]{
        \includegraphics[width=3.0in]{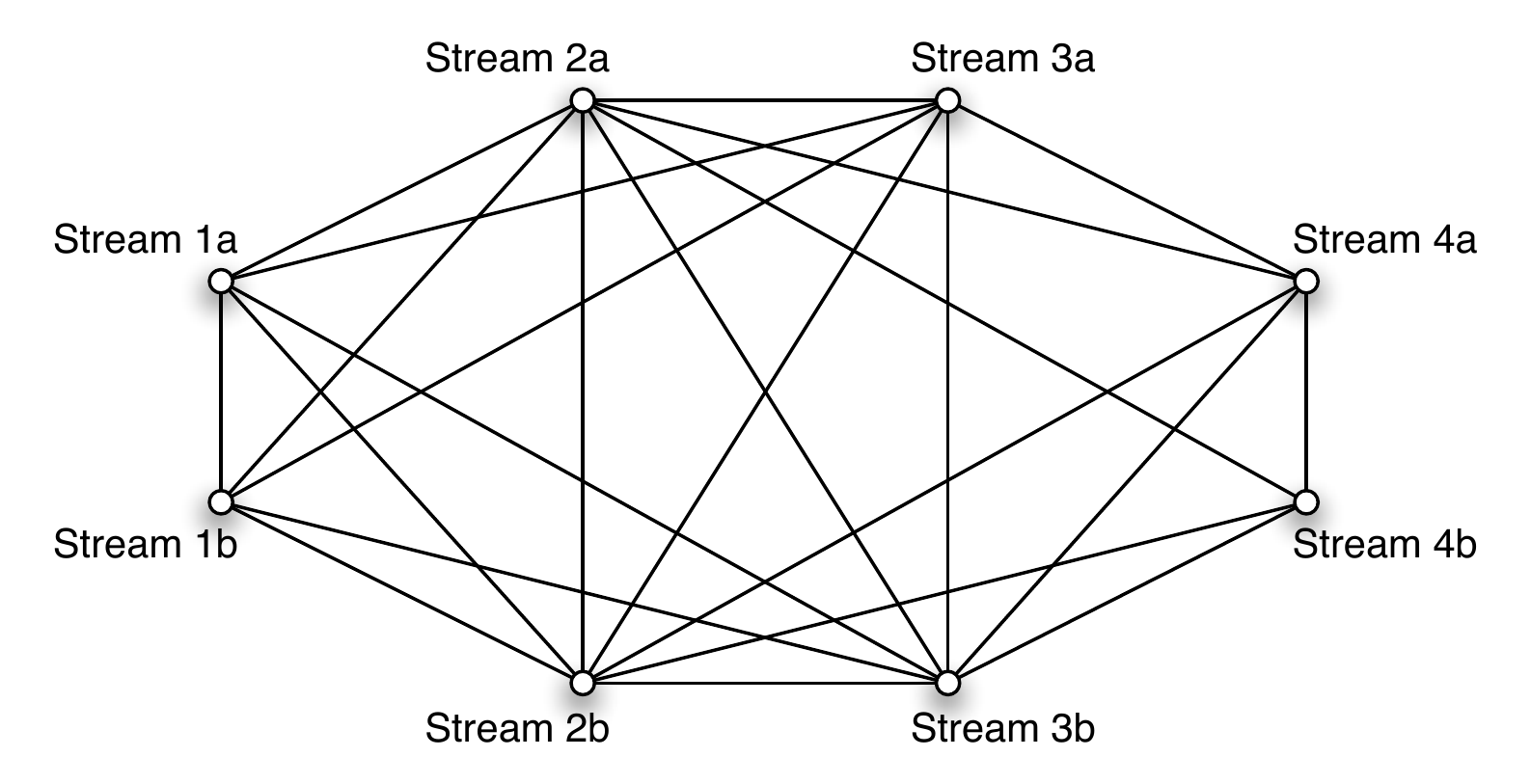}
        \label{fig:example-MIMO-graph}
        }
    \caption{Example illustrating (a) A simple network of MIMO nodes,
      and (b) the corresponding graph on which the distributed induced
      bounded-degree subgraph algorithm is run.}
\end{center}
\end{figure}

Consider a MIMO network in which each node possesses $q$ antenna
elements (therefore, $q$ degrees of freedom) and is capable of
transmitting on $q$ different streams to a neighboring node. Consider
a graph $G = (V,E)$ in which each vertex represents a potential stream
in a MIMO link. Two streams that interfere with each other are
connected by an edge in the graph. 
Thus, the streams comprising a single MIMO link
make up a $q$-vertex clique. 
Fig.\,\,\ref{fig:example-MIMO-network} shows
a simple MIMO network with five physical nodes. Assuming two streams
per MIMO link, Fig.\,\,\ref{fig:example-MIMO-graph} shows the
corresponding graph representing the interference pattern between the
different streams. 
Our goal is to maximize the number of active streams while ensuring that there are no more than
a certain number, say $k$, of streams interfering with any given
active stream. For example, in the case in which it takes exactly one degree of freedom at an
active receiver to suppress an undesired
interfering stream (i.e., no flexible interference suppression), an
active stream should have no more than $k=q-1$ other streams interfering
at the intended receiver. The problem now becomes 
that of finding the largest subset of vertices $V^\prime \subset
V$ such that for any vertex $v \in V^\prime$, there exist no more than
$k$ other vertices in $V^\prime$ connected to $v$ by an edge in
$G$. Thus, this becomes the maximum induced bounded-degree-$k$
subgraph (IBDS$(k)$) problem which is known to be NP-complete for general
graphs \cite{Tel1994}, although linear time solutions have been
discovered for certain types of graphs \cite{BodFlu1996}. 
The well-known maximum independent set problem is an instance of
maximum IBDS($k$) for $k=0$. We hasten to add that maximum IBDS($k$) is
different from another better-known and well-studied generalization of the maximum
independent set problem, the $k$-independent set problem (related to
the $k$-coloring of graphs) \cite{West2001}.  

We define a {\em maximal} IBDS($k$) as an induced bounded-degree-$k$
subgraph to which no additional
vertex can be added. We also define two simple variations of the
maximal IBDS($k$) problem for relevance in practical MIMO networks. Streams
corresponding to the same MIMO link (i.e., between the same pair of
nodes) are said to belong to the same {\em family}. Two
potential streams with at least one physical node in common are said
to belong to the same {\em superfamily}. In MIMO networks, practical
considerations often impose that a node does not
use two streams corresponding to different MIMO links
simultaneously. We define the maximal IBDS-R$(k)$ problem as 
the maximal IBDS($k$) problem with the additional restriction that for
any two vertices $v_1, v_2 \in V^\prime$, if $v_1$ and $v_2$ belong to
the same superfamily, then they belong to the same family. A second
variation is motivated by the fact that exploiting flexible
interference suppression requires that each MIMO link operate only a
subset (say, no more than $g$) of the maximum number of streams
possible and employ the remaining available degrees of freedom for interference
suppression. We define the maximal IBDS-R$(k,g)$ problem as the 
maximal IBDS-R$(k)$ problem with the restriction that no more than
$g$ vertices from the same family are selected into the maximal 
subgraph. 

\section{The Distributed Algorithm}
\label{sec:algo}

Fig.\,\,\ref{algo} presents a pseudo-code description of one round of
DML-IBDS($k$), a new distributed randomized algorithm for the
maximal IBDS($k$) problem (DML stands for distributed maximal). Each vertex $v$ holds four variables,
$r(v)$, $b(v)$, $\hat{b}(v)$ and $a(v)$. In the beginning before the
first round, each vertex $v$ computes a unique random 
integer $r(v)$. As in most distributed algorithms for networks, we
assume that the vertices possess unique ids and therefore, the
uniqueness of the random numbers is readily ensured by appending the
vertex id to the generated pseudo-random number. $r(v)$ is updated in
subsequent rounds and is not necessarily a unique number except at the
beginning of the first round. $b(v)$ is initially set to 0 and  
indicates whether or not vertex $v$, in some round, has $r(v)$ smaller
than $r(u)$ for all neighbors $u$ (this is what qualifies vertex $v$
for inclusion in the subgraph). $\hat{b}$ is initially set to
$-1$. When $\hat{b}$ is 1, it indicates that the vertex is chosen for
inclusion in the subgraph and when it is 0, it indicates that the
vertex has been eliminated from consideration. $a(v)$ is initially set
to $k+1$ and represents the maximum number of additional vertices that may be chosen for the
subgraph out of the set comprising of $v$ and its neighbors.

\begin{figure}[!t]
\begin{center}
\framebox[10cm]{
\parbox{7cm}{
{\small
\begin{tabbing}
wwi \= wi \= wi \= wi \= wi \= wi \= wi \= wi \= wi \kill
\mbox{~}\\ 
\mbox{~}01: \> {\bf Initialization} (at vertex $v$):\\
\mbox{~}02: \> \> $r(v) = $ pseudo-random integer\\
\mbox{~}03: \> \> $b(v) = 0$\\
\mbox{~}04: \> \> $\hat{b}(v) = -1$\\
\mbox{~}05: \> \> $a(v) = k+1$\\~\\
\mbox{~}06: \> {\bf One round of DML-IBDS($k$)} (at vertex $v$):\\
\mbox{~}07: \> \> if $a(v) > 0$:\\
\mbox{~}08: \> \> \> Send $r(v)$ to neighbors\\
\mbox{~}09: \> \> \> Receive $r(u)$ from each neighbor $u$\\
\mbox{~}10: \> \> \> if $r(v) \leq r(u)$ for every neighbor $u$:\\
\mbox{~}11: \> \> \> \> $b(v) = 1$\\
\mbox{~}12: \> \> \> Send $b(v)$ to neighbors\\
\mbox{~}13: \> \> \> Receive $b(u)$ from each neighbor $u$\\
\mbox{~}14: \> \> \> if $b(v) = 1$:\\
\mbox{~}15: \> \> \> \> if $\nexists u$ such that $b(u) = 0$ and $r(u) = r(v)$:\\
\mbox{~}16: \> \> \> \> \> $r(v) = \min\{ r(u) | r(u) > r(v) \}$\\
\mbox{~}17: \> \> \> if $\hat{b}(v) = -1$ and $b(v) = 1$:\\
\mbox{~}18: \> \> \> \> $\hat{b}(v) = 1$\\
\mbox{~}19: \> \> \> \> if $b(u) = 1$ for every neighbor $u$\\
\mbox{~}20: \> \> \> \> \> Halt\\
\mbox{~}21: \> \> \> \> Decrement $a(v)$\\
\mbox{~}22: \> \> \> \> Decrement $a(u)$ at each neighbor $u$\\
\mbox{~}23: \> \> if $a(v) \leq 0$:\\
\mbox{~}24: \> \> \> if $\hat{b}(v) = 1$:\\
\mbox{~}25: \> \> \> \> for each neighbor $u$:\\
\mbox{~}26: \> \> \> \> \> $a(u) = 0$ \\
\mbox{~}27: \> \> \> \> \> if $\hat{b}(u) = -1$\\
\mbox{~}28: \> \> \> \> \> \> $\hat{b}(u) = 0$\\
\mbox{~}29: \> \> \> Halt\\
\end{tabbing}
} } }
\end{center}
\caption{Pseudo-code description of DML-IBDS($k$). A vertex $v$
  is selected into the subgraph if $\hat{b} = 1$.} 
\label{algo}
\end{figure}

Each vertex $v$ sends its
random number, $r(v)$, to its neighbors and receives numbers from its
neighbors (lines 08--09). If none of $v$'s 
neighbors carries a smaller value of $r$, then vertex $v$ sets its
$b(v)$ to 1 (lines 10--11). Neighbors communicate their value of $b$ to each other
(lines 12--13). A vertex that has its $b$ value set to 1 will now consider
updating its $r$ value. If $v$ has no neighbors with the same $r$
value as itself or if all neighbors of $v$ with the same $r$ value are
already chosen for inclusion into the subgraph (i.e., $b=1$), then $v$ updates
its $r$ value to the smallest among the received values but larger
than $r(v)$ (lines 15--16). This step ensures that no more than one
of the neighbors of a vertex with $b=1$ will be selected into the
subgraph in the same round (thus guaranteeing that $a(v)$ does not
suddenly reach below zero; a vertex $v$ with $b(v)=1$ that does not yet
have $a(v) = 0$ will always possess an $r(v)$ equal to the smallest of
its neighbors, and therefore, only its smallest neighbor will
potentially be chosen into the subgraph.) 

\theorem The DML-IBDS($k$) algorithm finds a maximal subgraph
with degree bound $k$ in $O(n)$ rounds where $n$ is the number of
vertices in the graph. 

\begin{proof}
At the beginning of a round, consider the vertex $v$ with the
smallest value of $r(v)$ such that $\hat{b}(v) = -1$. By lines 15--16, any
vertex $u$ with $\hat{b}(u) = 1$ that is still participating in the
rounds will have $r(u)$ no smaller than $r(v)$. Therefore, vertex $v$
will have $\hat{b}$ set to 1 by the end of the round. Thus, in each
round, at least one additional vertex has its $\hat{b}$ set to one, and
therefore, the algorithm completes in $O(n)$ rounds.

A vertex $v$ reduces $a(v)$ by one whenever itself or one of its
neighbors is chosen into the subgraph (lines 21--22). However, $a(v)$ is never allowed to go
below zero for vertices that have $\hat{b}$ set to 1 (i.e., vertices
corresponding to streams chosen for the subgraph). Since a vertex
halts when $a(v) = 0$ (line 29), the induced 
subgraph will satisfy the desired degree bound. Also, it is a maximal
subgraph because a vertex halts only when $a(v) \leq 0$ (line 29) or when the
vertex along with all its neighbors has been chosen for inclusion into
the subgraph (line 20).
\end{proof}

In real networks, after a topology control algorithm such as RNG has
been executed, each physical node would have a bounded number of MIMO
links. The corresponding graph of interference pattern among the
streams, therefore, has a degree bound as well. The probability that any
given vertex $v$ ends up with an $r(v)$ no smaller than that of
any of its neighbors is less than or equal to $1/(1+deg_G(v))$. Given
a bound on the degree, each node has a probability greater than some
constant of having its $a(v)$ reduced by one. Since each of the $n$
vertices begins with $a=k$ degrees of freedom, even though the worst-case
number of rounds is $O(n)$, the actual number of rounds is $O(\log
nk)$ with high probability (i.e., with probability greater than or
equal to $1 - 1/(nk)^c$ for any $n \geq n_0$, $k > k_0$ for some positive
constants $c, k_0$ and $n_0$). 

Distributed solutions to variations of
maximal IBDS($k$) described in the previous section are only
trivially different and are denoted in this report
by DML-IBDS-R$(k)$ and DML-IBDS-R$(k,g)$.

\section{Results and Concluding Remarks}
\label{sec:results}

\begin{figure}[!t]
\begin{center}
\includegraphics[width=5.0in]{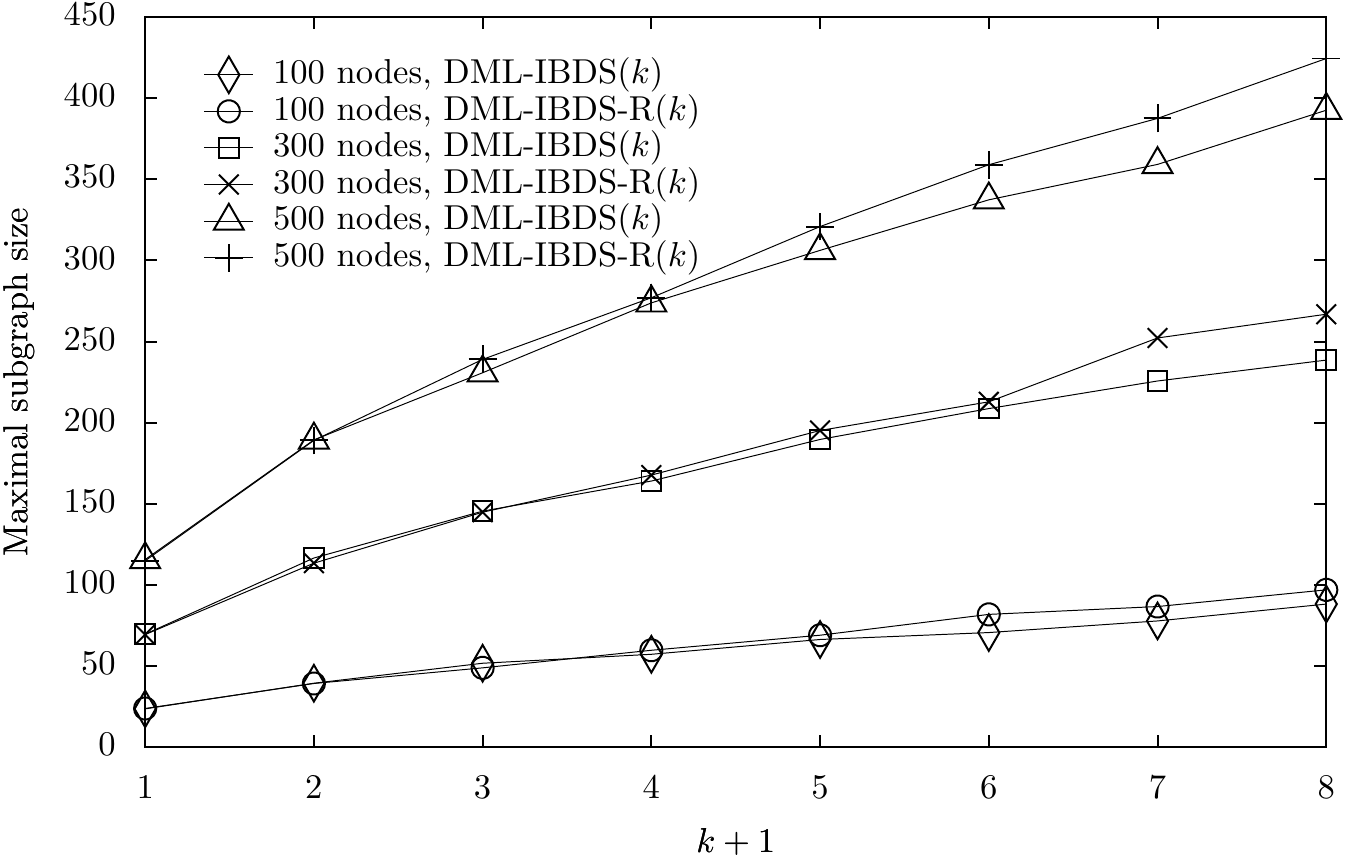}
\caption{Size of the subgraph with DML-IBDS($k$) and
  DML-IBDS-R$(k)$ plotted against $k+1$. Note that in the 
  absence of flexible interference suppression, $k+1$ may be thought
  of as the number of antenna elements per node.} 
\label{fig:numMatches}
\end{center}
\end{figure}

Fig.\,\,\ref{fig:numMatches} presents simulation results on the performance of
DML-IBDS($k$) and DML-IBDS-R$(k)$ as measured by the number of
vertices in the degree-bounded subgraph that is generated. The networks used in 
our experiments consist of nodes located randomly in a unit square
area and with the IBDS algorithms applied after the application of the
DRNG (Directed Relative Neighborhood Graph) topology control algorithm
\cite{LiHou2005-1313}. We choose the DRNG protocol because it is one
that can be employed in a multipath environment where a node cannot
readily deduce the distance and direction of its neighbors. Each data
point in this report represents an average of over 25 different
randomly generated networks.

As can be seen from Fig.\,\,\ref{fig:numMatches}, DML-IBDS-R$(k)$ performs
slightly better than that for the unrestricted version, DML-IBDS($k$), indicating
that the negative impact of receiver overloading is reduced when the active streams at a node
all belong to the same MIMO link. This suggests that the synchronization necessary to
allow simultaneous use of streams to different neighbors at a node may
not yield a higher aggregate throughput unless the stream control
gains are high (i.e., if the strongest channel modes are significantly
stronger than the weaker channel modes).

Fig.\,\,\ref{fig:numMatches} also shows that the number of streams used
(size of induced subgraph) does not increase linearly as the number of antenna elements increases. This is
obviously sub-optimal to a trivial solution that involves no stream
control: generate a maximal independent set, such as by running
DML-IBDS(0), and at all the MIMO links chosen to transmit on a 
stream just 
allow all the streams to transmit. Since practical MAC algorithms
do not usually approach the capacity corresponding to a maximal
independent set, this sublinear increase in the number of active
streams does not necessarily suggest that stream control is not
desirable in practice. In fact, as shown in \cite{SunSiv2004}, stream
control does improve performance in real contexts. Our results,
however, shed light on the 
significant negative effect of receiver overloading and help us
quantify the desired stream gains needed to offset the effect. For example,
in a 500-node network in the absence of flexible interference
suppression, the number of streams with one antenna element per node is
115 (subgraph size yielded by DML-IBDS-R$(0)$) while that with two antenna
elements is 189 (subgraph size yielded by DML-IBDS-R$(1)$). This implies  
that for stream control to increase capacity in a network with two
antenna elements per node, in the absence of
flexible interference suppression, the average gain due to stream
control should be about 22\% (2$\times$$115/189 - 1$).

\begin{figure}[!t]
\begin{center}
\includegraphics[width=5.0in]{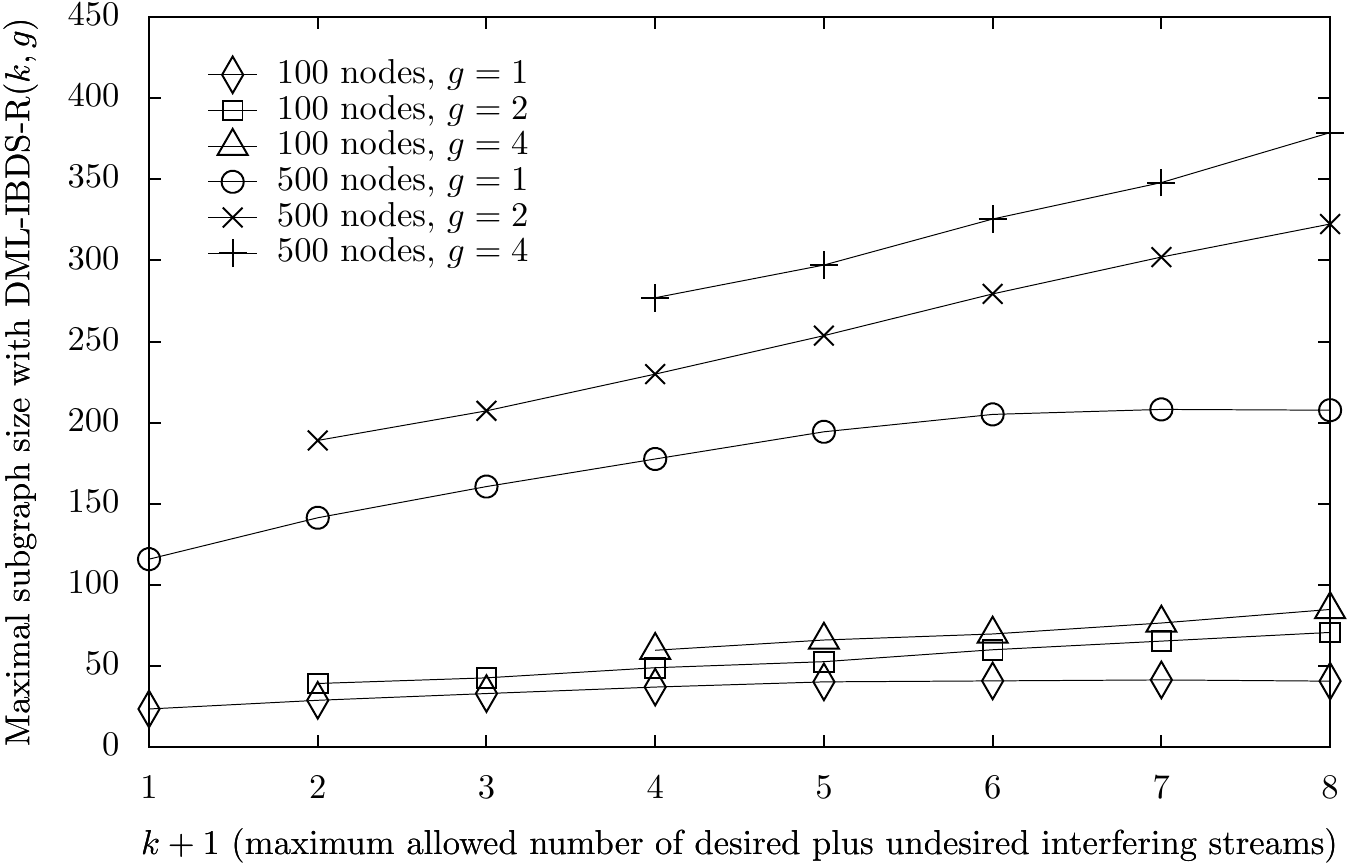}
\caption{Size of subgraph produced by DML-IBDS-R$(k,g)$, i.e., when
  each MIMO link is allowed no more than $g$ active streams and no
  more than $k+1$ total streams (desirable plus undesirable) interfere
  at any receiver.}
\label{fig:flexInterf}
\end{center}
\end{figure}

Fig.\,\,\ref{fig:flexInterf} shows us the gains achievable through
flexible interference suppression and permit a quantitative comparison to the
negative impact of receiver overloading. The subgraph size with
DML-IBDS-R$(k,g)$ indicates the number of active streams in the
network when the number of active streams per MIMO link is bounded by
$g$ and the total number of streams (both desirable and the undesirable
interfering ones) arriving at a receiver is bounded by $k+1$. For example, for 500-node networks,
Figs.\,\,\ref{fig:numMatches} and \ref{fig:flexInterf} show that the
subgraph size obtained with DML-IBDS-R$(1)$ 
lies somewhere between those obtained by DML-IBDS-R$(4,1)$ and
DML-IBDS-R$(5,1)$. If we have two antenna elements per node and if only
one stream ($g=1$) is used per MIMO link in order to exploit flexible
interference suppression, the remaining one available degree of freedom has to be 
sufficient to suppress up to five additional undesirable streams to exceed
the subgraph size that can be obtained by allowing both streams to
operate whenever possible. These results suggest that, unless each
available degree of freedom can be used to suppress several undesirable
interfering streams, it is better to use any available degrees of
freedom to active streams instead of flexible interference suppression.
These insights offered by our algorithmic results can be used to design
better MAC protocols for MIMO networks.

~\\
\noindent {\bf Acknowledgments}

The author wishes to thank Kapil R. Dandekar for introducing him to MIMO networks.
This work was supported in part by NSF Award CNS-0322797.

\bibliographystyle{IEEEtran}
\bibliography{ibds-mimo-report}

\begin{thebibliography}{10}
\providecommand{\url}[1]{#1}
\csname url@samestyle\endcsname
\providecommand{\newblock}{\relax}
\providecommand{\bibinfo}[2]{#2}
\providecommand{\BIBentrySTDinterwordspacing}{\spaceskip=0pt\relax}
\providecommand{\BIBentryALTinterwordstretchfactor}{4}
\providecommand{\BIBentryALTinterwordspacing}{\spaceskip=\fontdimen2\font plus
\BIBentryALTinterwordstretchfactor\fontdimen3\font minus
  \fontdimen4\font\relax}
\providecommand{\BIBforeignlanguage}[2]{{%
\expandafter\ifx\csname l@#1\endcsname\relax
\typeout{** WARNING: IEEEtran.bst: No hyphenation pattern has been}%
\typeout{** loaded for the language `#1'. Using the pattern for}%
\typeout{** the default language instead.}%
\else
\language=\csname l@#1\endcsname
\fi
#2}}
\providecommand{\BIBdecl}{\relax}
\BIBdecl

\bibitem{Pel2000}
D.~Peleg, \emph{Distributed Computing: A Locality-Sensitive Approach}.\hskip
  1em plus 0.5em minus 0.4em\relax SIAM, 2000.

\bibitem{BalBar2004}
H.~Balakrishnan, C.~L. Barrett, V.~S.~A. Kumar, M.~V. Marathe, and S.~Thite,
  ``The distance-2 matching problem and its relationship to the {MAC}-layer
  capacity of ad hoc wireless network,'' \emph{IEEE Journal on Selected Areas
  in Communications}, vol.~22, no.~6, pp. 1069--1079, August 2004.

\bibitem{HayMoh2005}
S.~Haykin and M.~Moher, \emph{Modern Wireless Communications}.\hskip 1em plus
  0.5em minus 0.4em\relax Prentice Hall, 2005.

\bibitem{DemIng2003}
M.~F. Demirkol and M.~A. Ingram, ``Stream control in networks with interfering
  {MIMO} links,'' in \emph{Proceedings of the IEEE Wireless Communications and
  Networking Conference}, 2003, pp. 343--348.

\bibitem{BluWin2002}
R.~S. Blum, J.~H. Winters, and N.~R. Sollenberger, ``On the capacity of
  cellular systems with {MIMO},'' \emph{IEEE Communication Letters}, vol.~6,
  no.~6, pp. 242--244, June 2002.

\bibitem{SunSiv2004}
K.~Sundaresan, R.~Sivakumar, M.~A. Ingram, and T.-Y. Chang, ``Medium access
  control in ad hoc networks with {MIMO} links: Optimization considerations and
  algorithms,'' \emph{IEEE Transactions on Mobile Computing}, vol.~3, no.~4,
  pp. 350--365, October-December 2004.

\bibitem{Tel1994}
J.~A. Telle, ``Complexity of domination-type problems in graphs,'' \emph{Nordic
  Journal of Computing}, vol.~1, no.~1, pp. 157--171, Spring 1994.

\bibitem{BodFlu1996}
H.~L. Bodlaender and B.~de~Fluiter, ``Reduction algorithms for constructing
  solutions in graphs with small treewidth,'' in \emph{Computing and
  Combinatorics}, ser. Lecture Notes in Computer Science, vol. 1090.\hskip 1em
  plus 0.5em minus 0.4em\relax Springer, 1996, pp. 199--208.

\bibitem{West2001}
D.~B. West, \emph{Introduction to Graph Theory}, 2nd~ed.\hskip 1em plus 0.5em
  minus 0.4em\relax Prentice Hall, 2001.

\bibitem{LiHou2005-1313}
N.~Li and J.~C. Hou, ``Localized topology control algorithms for heterogeneous
  wireless networks,'' \emph{IEEE/ACM Transactions on Networking}, vol.~13,
  no.~6, pp. 1313--1324, December 2005.

\end{thebibliography}

\end{document}